\documentclass[fleqn]{2017SCGE}
\usepackage{multirow}
\usepackage{lineno}
\begin{document}
\ensubject{subject}
\ArticleType{Article}
\SpecialTopic{SPECIAL TOPIC:}
\Year{2019}
\Month{June}
\Vol{00}
\No{0}
\DOI{10.1007/000000}
\ArtNo{0000000}
\ReceiveDate{xxx xx,2019}
\AcceptDate{xxx xx, 2019}

\title{An Improved Evaluation of the Neutron Background in the PandaX-II Experiment}{An Improved Evaluation of the Neutron Background in the PandaX-II Experiment}
\author[1,13]{Qiuhong Wang\footnote{Corresponding author: wangqiuhong@sinap.ac.cn}}{}
\author[2]{Abdusalam Abdukerim}{}
\author[2]{Wei Chen}{}
\author[2]{Xun Chen}{}
\author[3]{Yunhua Chen}{}
\author[2]{Xiangyi Cui}{}
\author[4]{\\Yingjie Fan}{}
\author[1]{Deqing Fang}{}
\author[2]{Changbo Fu}{}
\author[5]{Lisheng Geng}{}
\author[2]{Karl Giboni}{}
\author[2]{Franco Giuliani}{}
\author[2]{Linhui Gu}{}
\author[3]{\\Xuyuan Guo}{}
\author[2]{Ke Han}{}
\author[2]{Changda He}{}
\author[2]{Di Huang}{}
\author[3]{Yan Huang}{}
\author[6]{Yanlin Huang}{}
\author[2]{Zhou Huang}{}
\author[4]{\\Peng Ji}{}
\author[2,7]{Xiangdong Ji}{}
\author[8]{Yonglin Ju}{}
\author[2]{Yihui Lai}{}
\author[2]{Kun Liang}{}
\author[8]{Huaxuan Liu}{}
\author[2,7]{Jianglai Liu\footnote{Spokesperson: jianglai.liu@sjtu.edu.cn}}{}
\author[2]{\\Wenbo Ma}{}
\author[1]{Yugang Ma}{}
\author[9]{Yajun Mao}{}
\author[2]{Yue Meng}{}
\author[2]{Parinya Namwongsa}{}
\author[2]{Kaixiang Ni}{}
\author[3]{\\Jinhua Ning}{}
\author[2]{Xuyang Ning}{}
\author[8,2]{Xiangxiang Ren}{}
\author[3]{Changsong Shang}{}
\author[2]{Lin Si}{}
\author[10]{Andi Tan}{}
\author[11]{\\Anqing Wang}{}
\author[1]{Hongwei Wang}{}
\author[11]{Meng Wang}{}
\author[9]{Siguang Wang}{}
\author[8]{Xiuli Wang}{}
\author[10,12,2]{Zhou Wang}{}
\author[4]{\\Mengmeng Wu}{}
\author[3]{Shiyong Wu}{}
\author[2]{Jingkai Xia}{}
\author[10,12]{Mengjiao Xiao}{}
\author[7]{Pengwei Xie}{}
\author[11]{Binbin Yan}{}
\author[2]{\\Jijun Yang}{}
\author[2]{Yong Yang}{}
\author[4]{Chunxu Yu}{}
\author[11]{Jumin Yuan}{}
\author[10]{Dan Zhang}{}
\author[2]{Hongguang Zhang}{}
\author[2]{\\Tao Zhang}{}
\author[2]{Li Zhao}{}
\author[6]{Qibin Zheng}{}
\author[3]{Jifang Zhou}{}
\author[2]{Ning Zhou}{}
\author[5]{Xiaopeng Zhou}{}

\AuthorMark{Wang Q H} 
\AuthorCitation{Wang Q H, et al}

\address[ ]{(PandaX-II Collaboration)}

\address[1]{Shanghai Institute of Applied Physics,
  Chinese Academy of Sciences, Shanghai 201800, China}
\address[2]{INPAC and School of Physics and Astronomy, Shanghai Jiao Tong University,\\ 
  Shanghai Laboratory for Particle Physics and Cosmology, Shanghai 200240, China}
\address[3]{Yalong River Hydropower Development Company, Ltd., 
  288 Shuanglin Road, Chengdu 610051, China} 
\address[4]{School of Physics, Nankai University, Tianjin 300071, China}
\address[5]{School of Physics \& International Research Center for Nuclei and Particles in the Cosmos \&\\Beijing Key Laboratory of Advanced Nuclear Materials and Physics, Beihang University, Beijing 100191, China}
\address[6]{School of Medical Instrument and Food Engineering, University of Shanghai for Science and Technology, Shanghai 200093, China}
\address[7]{Tsung-Dao Lee Institute, Shanghai 200240, China}
\address[8]{School of Mechanical Engineering, Shanghai Jiao Tong
  University, Shanghai 200240, China} 
\address[9]{School of Physics, Peking University, Beijing
  100871,China} 
\address[10]{Department of Physics, University of
  Maryland, College Park, Maryland 20742, USA} 
\address[11]{School of Physics
  and Key Laboratory of Particle Physics and Particle Irradiation
  (MOE), Shandong University, Jinan 250100, China}
\address[12]{Center of High Energy Physics, Peking
  University, Beijing 100871, China}
\address[13]{University of Chinese Academy of Sciences, Beijing 100049, China}

\abstract{In dark matter direct detection experiments,
   neutron is a serious source of background, which can mimic 
   the dark matter-nucleus scattering signals. In this
  paper, we present an improved evaluation of the neutron background
  in the PandaX-II dark matter experiment by a novel approach.
  Instead of fully relying on the Monte Carlo simulation,
  the overall neutron background is determined from the neutron-induced high energy signals in the data.
  In addition,
  the probability of producing a dark-matter-like background per neutron is evaluated with a complete Monte Carlo generator, where the correlated emission of neutron(s) and $\gamma$(s) in the ($\alpha$, n) reactions and spontaneous fissions is taken into consideration.   With this method, the neutron
  backgrounds in the Run 9 (26-ton-day) and Run 10 (28-ton-day) data
  sets of PandaX-II are estimated to be 0.66$\pm$0.24 and
  0.47$\pm$0.25 events, respectively.}


\maketitle



\begin{multicols}{2}

\section{Introduction}
\label{sec:introduction}
Modern astronomical and cosmological observations have firmly
established the existence of dark matter (DM) in our Galaxy and the
Universe~\cite{doi:10.1146/annurev.aa.25.090187.002233}.  Since the
seminal paper by Goodman and Witten~\cite{Goodman:1984dc}, there has
been an intensive global effort of DM direct detection,
searching for the nuclear recoil (NR) of the atomic
nucleus in the target induced by galactic DM particles
~\cite{Cushman:2013zza,Liu:nphys4039,Aprile:2009dv}. 

As one of such experiments, the PandaX experiment, located at 
the China Jinping Underground Laboratory~\cite{Yu-Cheng:2013iaa}
uses liquid xenon as the DM direct detection target~\cite{Cao:2014jsa,Xiao:2014xyn}.
The second phase of this experiment, PandaX-II~\cite{Tan:2016zwf}, 
contains a total of 1.1~tons of liquid xenon in a stainless steel (SS)
cryostat with low levels of radioactivity~\cite{Zhang:2016pgh}. Approximately 580~kg of liquid xenon is confined
inside the 60-cm high and 60-cm diametral dual phase xenon time
projection chamber (TPC) with Hamamatsu R11410 photomultipliers (PMTs)
located at the top and bottom. For each recoil event, the prompt scintillation photons ($S1$) and the delayed
electroluminescence photons ($S2$) from ionized electrons can be
detected by those two arrays of PMTs. Highly reflective
polytetrafluoroethylene (PTFE) layers are installed surrounding the TPC 
to enhance the photon collection
efficiency.  So far, PandaX-II has released two data sets, the Run 9
(79.6 live days, 26-ton-day exposure)~\cite{Tan:2016zwf} and Run
10 (77.1 live days, 28-ton-day exposure)~\cite{Cui:2017nnn}.

From the background point of view, the charge ratio of $S2$ to $S1$ offers a
strong discriminant for the NR signals against the electron recoil
(ER) background events from $\gamma$s and $\beta$s. On the other hand,
the neutrons could produce single-site
scattering NR (SSNR) events that highly resemble the DM interactions.
External neutrons can be reduced to a negligible level via carefully
designed shielding. However, neutrons can be produced internally from
radioactivities in the detector materials through two main reactions,
the ($\alpha$, n) reaction and spontaneous fission (SF) of
$^{238}$U, and bring about a non-negligible background to the DM detection.

Traditionally, neutron background evaluation is heavily
simulation-based. In this paper, we present a novel data-driven
method to evaluate the neutron background in PandaX-II, utilizing
neutron-induced high energy gamma (HEG) events as the
normalization. The rest of this paper is organized as follows.
In \cref{sec:method}, we explain our improved method of
the SSNR background evaluation.
In \cref{sec:ambe_calibration}, we use the $^{241}$Am-$^{9}$Be (AmBe)
neutron calibration data to understand the SSNR and HEG events. In
\cref{sec:material_neutron}, we discuss our improved neutron
generator model and establish an expected ratio between the two
types of events originating from the detector
materials. A new estimation of the neutron background level in
PandaX-II is given in \cref{sec:pandax_nbkg}, followed by a
summary in \cref{sec:summary}.

\section{Method}
\label{sec:method}
Traditionally, the SSNR background estimation for DM experiments has
the following three ingredients, 1) radioactivities of detector
materials, 2) a model which converts the radioactivities to the number of neutrons
produced and their energy spectrum, and 3) a detector Monte Carlo
(MC) simulation which uses the model as the event generator and carries out
detailed simulation of neutron interactions and the detector response.
The SSNR background is calculated as
\begin{eqnarray}
  \label{eq:ni}
  N_{\rm{ssnr}} &=& \displaystyle\sum_{ij} P_{{\rm ssnr},ij} \times n_{ij} \times T\,\nonumber\\
  n_{ij} &=& Y_{ij} \times A_{ij} \times M_i \,.
\end{eqnarray}
In these expressions, $n_{ij}$ is the neutron production rate from a
detector component $i$ due to a given radionuclide or a reaction chain
$j$, calculated by combining the radioactivity per unit mass $A_{ij}$, 
the neutron yield $Y_{ij}$, and the component mass
$M_i$. $Y_{ij}$ is typically computed by the SOURCES-4A
code~\cite{Wilson:1999so}, which calculates the ($\alpha$, n) and SF
neutron production rates and spectra using evaluated libraries. 
$P_{{\rm ssnr},ij}$ is the probability of producing an SSNR signal
per neutron calculated by the detector MC and $T$ is the
duration of the measurement.

The above method has two serious drawbacks. First, SOURCES-4A
and the resulting neutron generator treat every neutron as 
an isolated particle. This is, in fact, a biased assumption - there is a high probability that the neutron(s) is generated in association with $\gamma$(s) in the ($\alpha$, n) and SF reactions, leading to multi-site scatterings and mixed ER-NR energy depositions.
Second, the uncertainty in radioassay of detector materials $A_{ij}$ would be directly
translated into a normalization uncertainty in the SSNR rate, which
lacks experimental handle.

In light of the above, we developed a new method to estimate the
neutron background. A complete neutron generator was
developed which incorporates the correlated emission of
neutron(s) and $\gamma$(s). This allows us to establish a more robust
relationship between the SSNR events and the neutron-induced HEG events. Then the SSNR
background can be estimated as
\begin{eqnarray}
  \label{eq:heg_rmc}
  N_{\rm{ssnr}} = N_{\rm{heg}} \times R_{\rm{mc}} = N_{\rm{heg}} \times  \Bigg(\displaystyle\frac{\sum_{ij} P_{{\rm ssnr},ij} \times n_{ij}}{\sum_{ij} P_{{\rm heg},ij} \times n_{ij}}\Bigg)
\end{eqnarray}
where $N_{\rm{heg}}$ is the number of identified HEG events from DM data, which serves
as an {\it in situ} normalization. $R_{\rm{mc}}$ is an MC-based ratio
between the summed SSNR and HEG events from individual components, obtained
by folding the neutron production rate $n_{ij}$ and the probability to
produce an SSNR (HEG) event, $P_{{\rm ssnr},ij}$ ($P_{{\rm heg},ij}$).  In
what follows, we shall first use the AmBe calibration to understand
the connection between SSNR and HEG events, then discuss our evaluation of
$R_{\rm{mc}}$ and $N_{\rm{heg}}$ in turn.

\section{The AmBe calibration}
\label{sec:ambe_calibration}
\subsection{The AmBe neutron source}
\label{sec:ambe_neutron_source}
\begin{figure}[H]
  \centering
  \includegraphics[width=1.0\columnwidth]{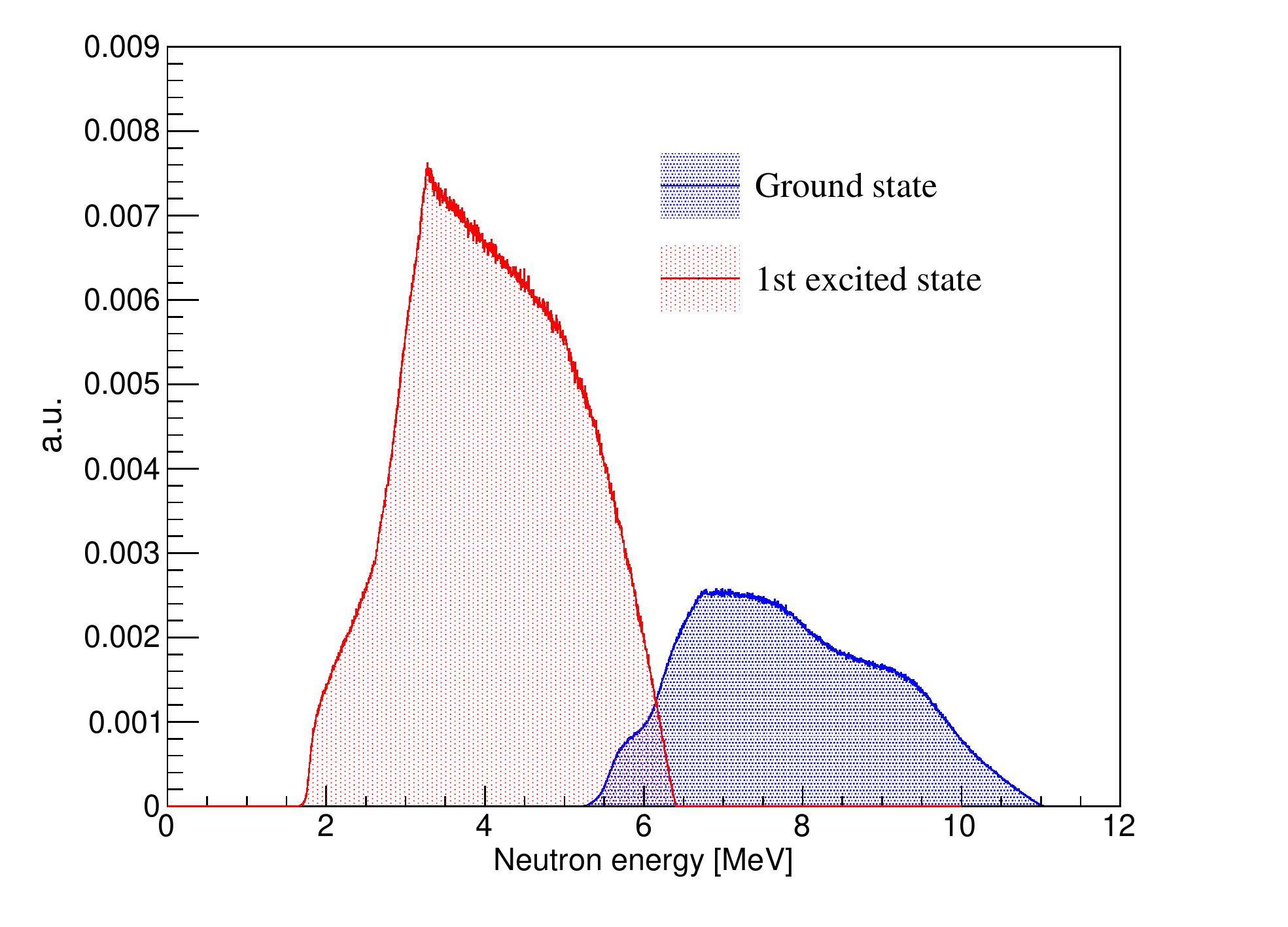}
  \caption{The neutron energy spectrum of the AmBe source from simulation, with the two histograms corresponding to ($\alpha$,n$_0$) (red) and ($\alpha$,n$_1$) (blue).}
  \label{fig:AmBe_spec}
\end{figure}

Two AmBe sources with the identical design were fabricated by the Atomic
High Technology Co., Ltd.~\footnote{For details, see \url{http://en.atom-hitech.com/product/51.html}.}. The sources produce neutrons via the ($\alpha$,n) reaction
\begin{eqnarray}
^{9}\rm{Be}+\alpha \to \rm{n}+^{12}\rm{C}^*\,.\nonumber
\end{eqnarray}
Two channels, ($\alpha$,n$_0$) and ($\alpha$,n$_1$), dominate the reaction, in which
$^{12}$C$^*$ is produced in the ground and the first
excited state, respectively. The latter channel emits a single
de-excitation $\gamma$ ray of 4.4~MeV together with the neutron. 
The neutron energy spectra of the two channels can be calculated by combining
a stand-alone Geant4-based~\cite{Agostinelli:2002hh,Allison:2006ve} simulation
(energy loss process of $\alpha$) and the JENDL database~\cite{jendl}
(neutron production) and are shown in \cref{fig:AmBe_spec}. 
One of the two AmBe sources was deployed into the
Daya Bay detector~\cite{dayabay}, where the simulated energy spectra were
validated through measurement~\cite{Gu:2018}.
The other one was used for the calibration of PandaX-II.

The AmBe source was deployed inside two separate horizontal PTFE tubes 
encircling the cryostat, at a vertical height of 1/4 (lower) and
3/4 (upper) of the TPC. The AmBe calibration runs in PandaX-II were 
performed more than 20 times, interspersed among the DM data taking,
with the source located at different locations in the loops.  In
total,  \linebreak \vspace{-8mm}

\noindent 6.8 and 22.8 days of AmBe data were taken in Run 9 and Run 10,
respectively, with a source-induced rate of approximately 2 Hz.  For
different source locations, no significant difference was identified
in the data or the MC, so we made no distinction in the calibration data analysis.

\subsection{AmBe data}
\label{sec:ambe_data}
AmBe neutrons can produce the SSNR signal through elastic
neutron-xenon-nucleus scattering if the scattering happens to be
single-site in the TPC. More often, however, AmBe neutrons would
scatter with xenon nuclei inelastically, producing NR
signals mixed with ER signals from the nuclear de-excitation. In
both cases above, the energy deposition is ``prompt'', and may be
further mixed with the ``prompt'' ER signals produced by the 4.4~MeV
$\gamma$ ray. On the other hand, after being thermalized, neutrons are
also likely to be captured by xenon nuclei, which produces ``delayed''
high energy $\gamma$ rays with a few $\mu$s delay time. These
neutron-induced HEGs, regardless if produced promptly or by delayed
capture, are related to the SSNR events through the parent neutrons.

In the AmBe calibration data sets, SSNR events are selected using
identical cuts for dark matte data
selection~\cite{Tan:2016zwf,Cui:2017nnn}, with $S1$ from 3 to 45
photoelectrons (PE) and $S2$ from 100 PE (raw) to 10000 PE
(corrected for position uniformity). These cuts correspond to
an electron-equivalent energy range between 1 and 10~keV$_{ee}$ approximately.
The same fiducial volume (FV) cuts in Run 9 and Run 10 are adopted here.

\begin{figure}[H]
  \centering
  \includegraphics[width=1.0\columnwidth]{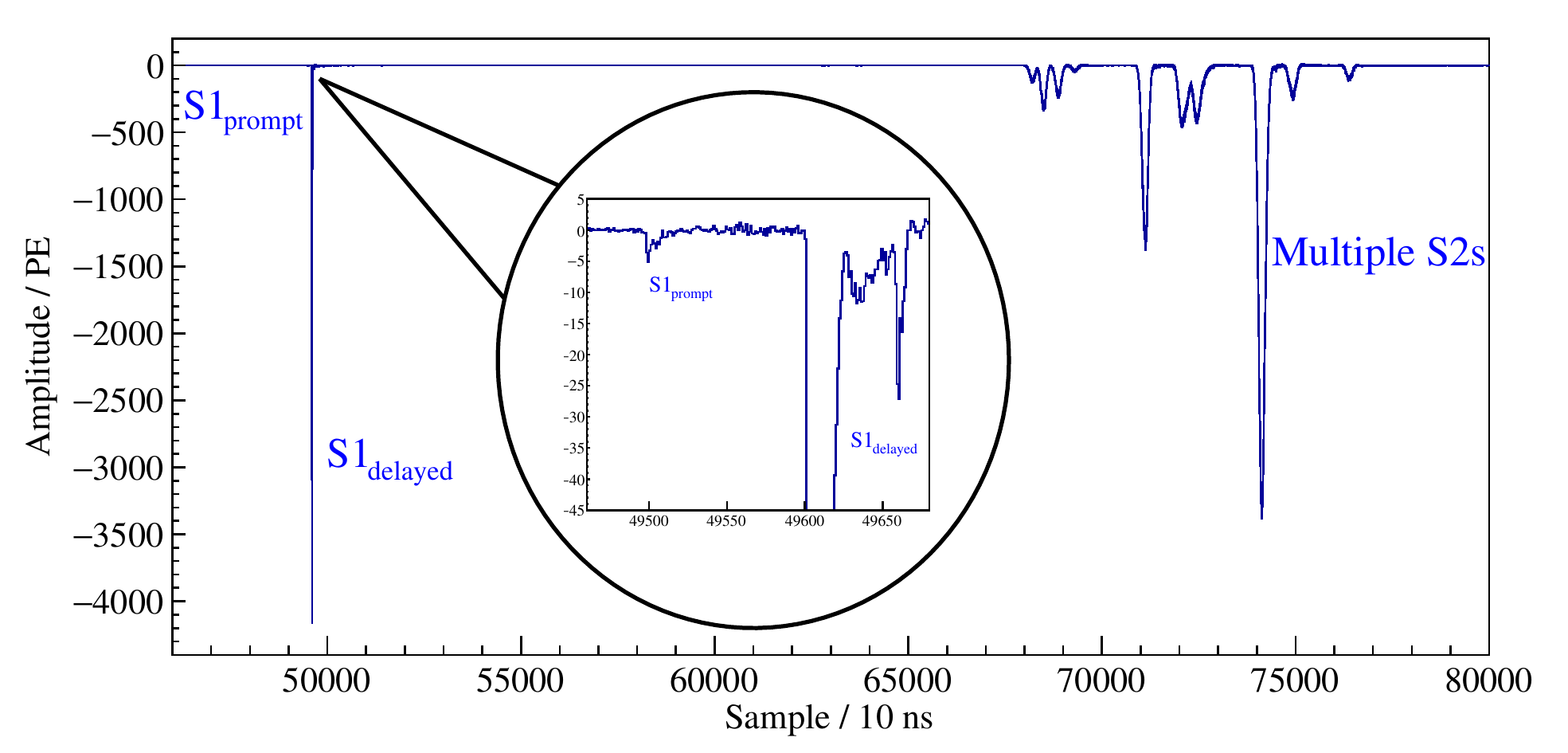}
  \caption{An example of waveforms of high energy neutron capture
    events in AmBe runs, in which the prompt NR occurred before the
    delayed HEGs.}
  \label{fig:high_e_waveform_ambe}
\end{figure}

The HEGs are easy to be identified because their total energy can be
significantly higher than those from natural radioactivities. In addition,
$\gamma$s can be produced at different times (prompt or delayed) and
tend to have multiple scatterings with xenon, resulting in multiple $S1$s and $S2$s.
A small fraction of them contain a separated NR signal before
the HEGs from the delayed capture. An example of such waveforms is
shown in \cref{fig:high_e_waveform_ambe}.

In the HEG event selection, the data quality cuts developed \linebreak \vspace{-8mm}

\noindent in Ref.~\cite{Tan:2016diz} are inherited for each $S1$ and $S2$.
For each scattering, the horizontal position is reconstructed
using the so-called PAF reconstruction~\cite{Tan:2016zwf}. The approximate
vertical positions of the scatterings are estimated by pairing these $S2$s with the maximum
$S1$~\footnote{Since neutron capture in xenon happens within a few
  $\mu$s (\cref{fig:time_separation_comparison}), it is reasonable
  to simply use $S1_{\rm max}$ as the $T_0$ of the drift.}.  Since we
are interested in the HEGs with energies larger than 5.0~MeV (Q value of $^{208}$Tl, the
maximum ER energy by natural radioactivity), there is no need of a
restrictive FV cut for background rejection. Also to avoid the bias in the horizontal position reconstruction caused by the PMT saturation for HEGs, we select all $S2$s within an extended FV (EFV) with a loose radius-square cut of
$r^2<1000$~cm$^2$ and a vertical position cut of 1.7 and 3.3~cm away
from the cathode and gate, respectively, resulting in a target xenon
mass of 502~kg. The total energy is computed by combining all $S1$s and
$S2$s within the EFV~\footnote{Note that in the data, the contributions to $S1$ from the energy depositions inside or outside the EFV cannot be separated. This leads to a slight discrepancy in the energy reconstruction in the data and MC, however its impact is constrained by the spectra comparison in \cref{fig:highe_spec_comparison}.}. To avoid saturation of the PMTs at this high
energy, the $S2$ signals in the bottom PMT array ($S2_{\rm b}$), scaled
by a factor of 3.5, are used in the reconstruction. 
A corrected energy, $E_{\rm{corr}}$, is obtained by further applying a polynomial correction to align the energy of the identified $\gamma$ events, i.e. 609 keV ($^{214}$Bi), 911/934 keV ($^{228}$Ac/$^{214}$Bi), 1173 keV ($^{60}$Co), 1332 keV ($^{60}$Co), 2614 keV ($^{208}$Tl), 4439 keV ($^{12}$C$^*$), 6467 keV ($^{132}$Xe$^*$) and 9255 keV ($^{130}$Xe$^*$), to their expected values.

\begin{figure}[H]
  \centering
  \includegraphics[width=1.0\columnwidth]{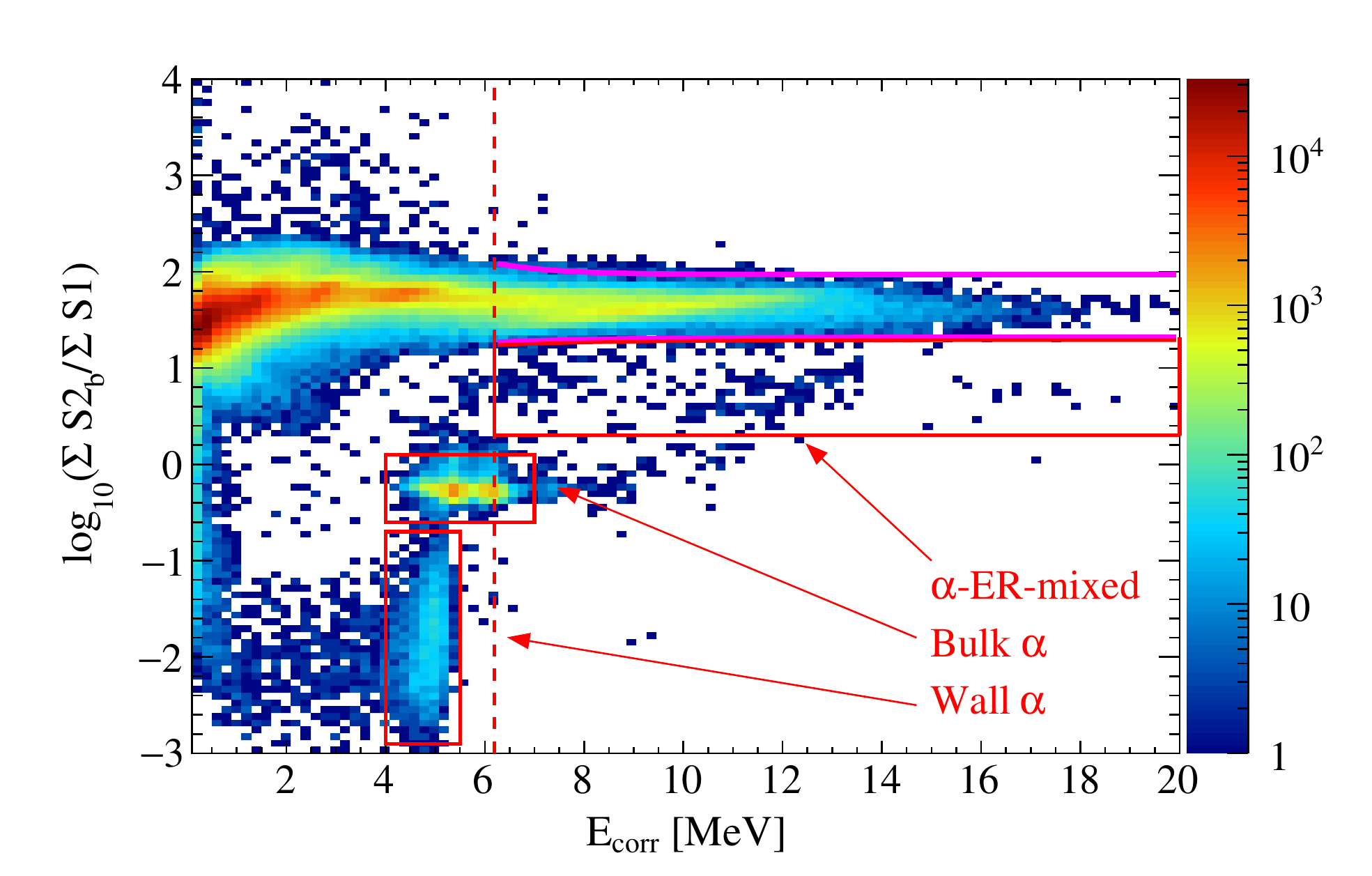}
  \includegraphics[width=1.0\columnwidth]{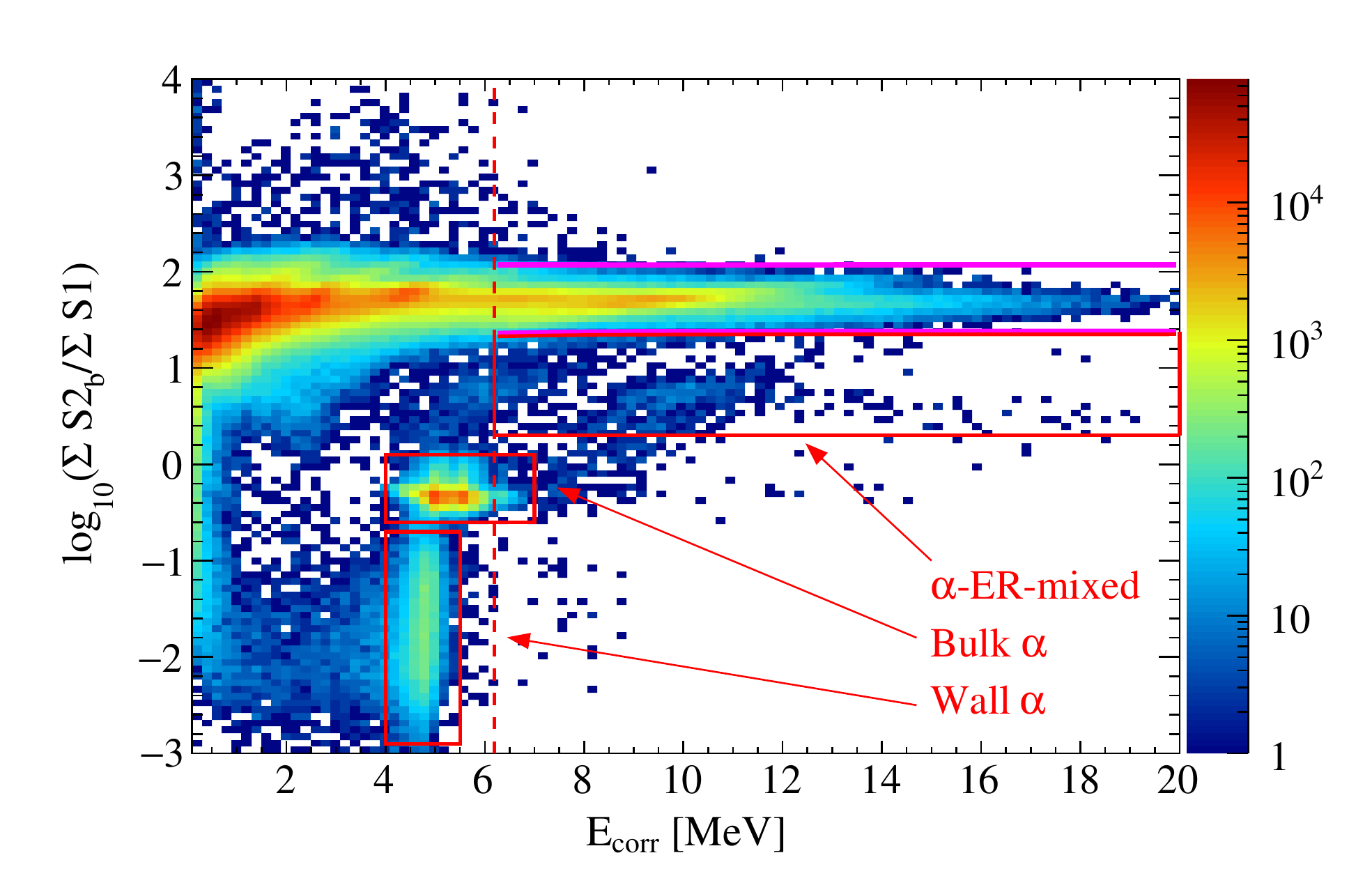}
  \caption{The distributions of events in $\log_{10}(\sum S2_{b}/\sum
    S1)$ vs. $E_{\rm corr}$ in the Run 9 (top) and Run 10 (bottom) AmBe
    data. The $\pm$3$\sigma$ cuts of the ER band are shown as the
    solid magenta lines. The dashed red lines represent the 6.2~MeV
    cut for the HEGs. Also indicated below the ER band are the
    $\alpha$-ER-mixed, bulk $\alpha$s, and wall $\alpha$ events
    near the PTFE reflectors.}
\label{fig:AmBe_energy_band}
\end{figure}

The event distributions in $\log_{10}(\sum S2_{b}/\sum S1)$ vs.
$E_{\rm{corr}}$ for the Run 9 and Run 10 AmBe data are shown in
\cref{fig:AmBe_energy_band}. HEGs are selected within a
3$\sigma$ cut on the high energy ER band, and with $E_{\rm corr}>$ 6.2~MeV,
about 3$\sigma$ away from 5~MeV, as indicated in
\cref{fig:AmBe_energy_band}.
Some events are observed with lower vertical values than
the main ER band. They are all $\alpha$-related events, resulting in
smaller ionization signals ($S2$s). We categorize the events into three classes according to the vertical values from high to low, i.e., the mixed $\alpha$-ER-events 
from the internal radon background, the bulk $\alpha$s, and the wall $\alpha$s
whose ionization signals cannot be efficiently
extracted. The $\alpha$-ER-mixed events have a rate much lower than the
AmBe-induced HEG rate here and therefore can be safely omitted.

The number of selected HEG and SSNR events and their ratios are summarized in
\cref{tab:AmBe_lowe_highe_ratio}, as compared with the simulation to be described in
\cref{sec:ambemc}.

\subsection{AmBe simulation}
\label{sec:ambemc}
A custom Geant4-based program (BambooMC), with the full PandaX-II
geometry implemented, is used to simulate energy depositions in the
detector by the radiation from the AmBe source. An event generator
that samples the neutron energy from \cref{fig:AmBe_spec}, with
correlated emission of 4.4~MeV $\gamma$ ray for the ($\alpha$,n$_1$)
channel, is used to create primary particles.  \linebreak \vspace{-8mm}

\begin{table}[H]
\footnotesize
\caption{The number of SSNR, HEG events and their ratios in Run 9 and Run 10 AmBe calibration data. The corresponding ratios from the simulation are also listed.}
\label{tab:AmBe_lowe_highe_ratio}
\doublerulesep 0.1pt \tabcolsep 10pt
\centering
\begin{tabular}{c|ccc|c}
\toprule
    	\multirow{2}*{\shortstack{AmBe Run}} & \multicolumn{3}{c|}{Data} & MC\\\cline{2-5}
      & \# SSNR & \# HEG & Ratio	& Ratio\\
    	\hline
	Run 9 & 3415 & 49159 & 1/14.4 & 1/14.7\\
	Run 10 & 10390 & 151783 & 1/14.6 & 1/15.2\\
\bottomrule
\end{tabular}
\end{table}

\noindent Note that the correlated $\gamma$ emission in this generator is implemented by hand, effectively equivalent to the improved ($\alpha$, n) generator discussed later in \cref{sec:neutron_alpha_n}. For SSNR events, the energy depositions
are converted to simulated $S1$ and $S2$ signals using a NEST-based
model~\cite{Lenardo:2014cva}. For HEG events, instead of running the
NEST-based simulation to produce $S1$s and $S2$s, we simply sum the
true energy depositions within the EFV, and make a gentle energy-dependent
Gaussian smearing with a $\sigma$ varying from 4\% to 10\% from 1~MeV to 10~MeV,
to best match the measured HEG spectrum from the AmBe data.
The efficiencies of event detection and selection cuts from AmBe data
are also incorporated into the MC simulation for the SSNR and HEG events.

\begin{figure}[H]
\centering
\includegraphics[width=1.0\columnwidth]{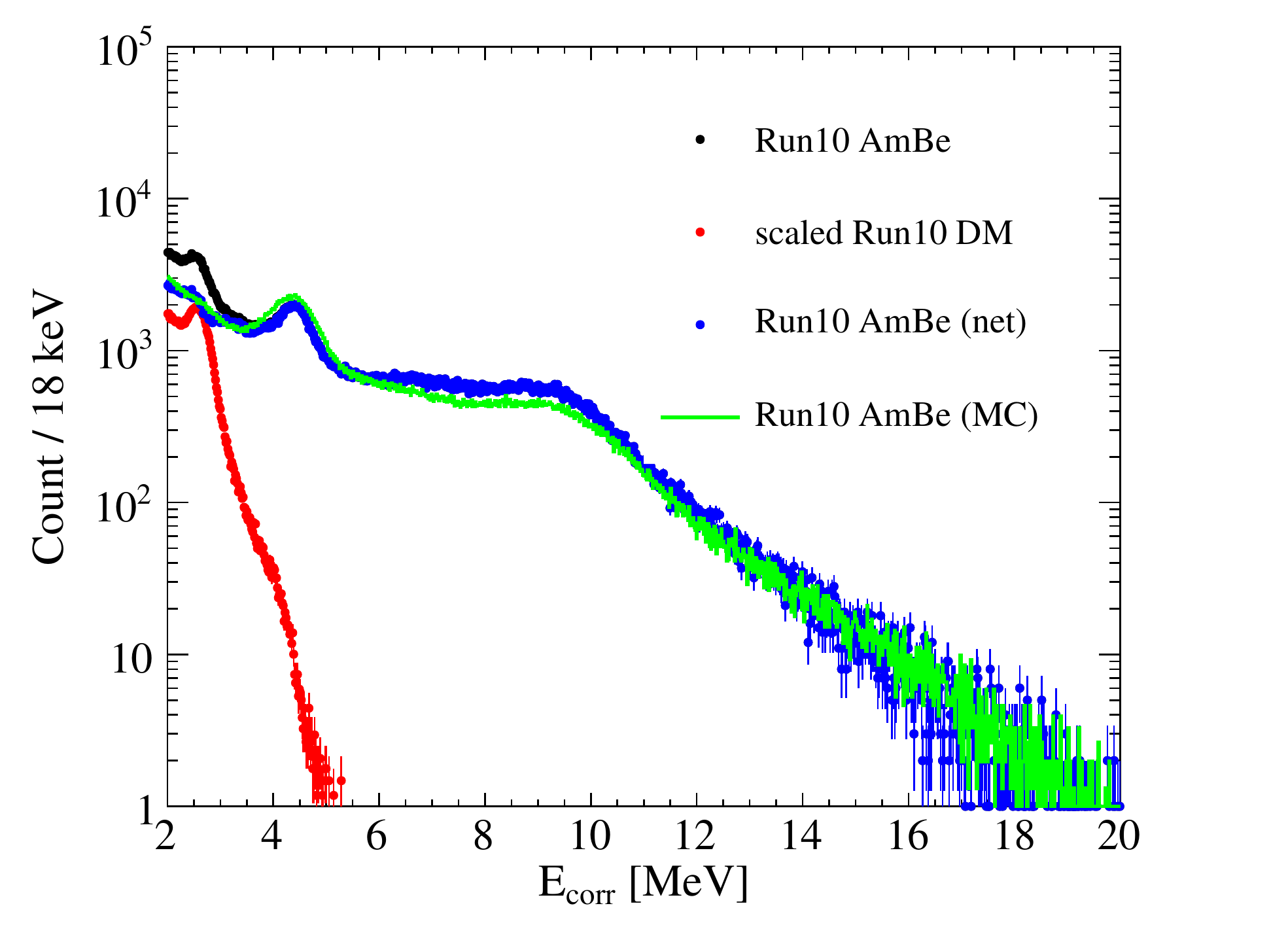}
\caption{The raw AmBe (black), background (red), and net AmBe (blue) energy spectra in Run 10. The background spectrum is obtained from the DM data, scaled by the run durations. The net AmBe spectrum is obtained by subtracting the background from the raw AmBe spectrum. The simulated energy spectrum is overlaid in green.}
\label{fig:highe_spec_comparison}
\end{figure}

The MC simulation is validated in several ways. In
\cref{fig:highe_spec_comparison}, the spectra of the measured
(non-AmBe background subtracted) and simulated $E_{\rm{corr}}$
are overlaid, where the simulation is normalized to the data between 2 and 20 MeV.
The spectra agree with each other well.
A residual difference between Runs 9 and 10 is observed, which is likely from the difference in the PMT saturation because of the varying supply voltages~\cite{Cui:2017nnn}. The fractional difference (data$-$MC) of the HEGs between 6.2 to 20~MeV is
16.3\% for Run 9 and 18.3\% for Run 10, indicating the level of 
systematic uncertainties.

\begin{figure}[H]
  \centering
  \includegraphics[width=1.0\columnwidth]{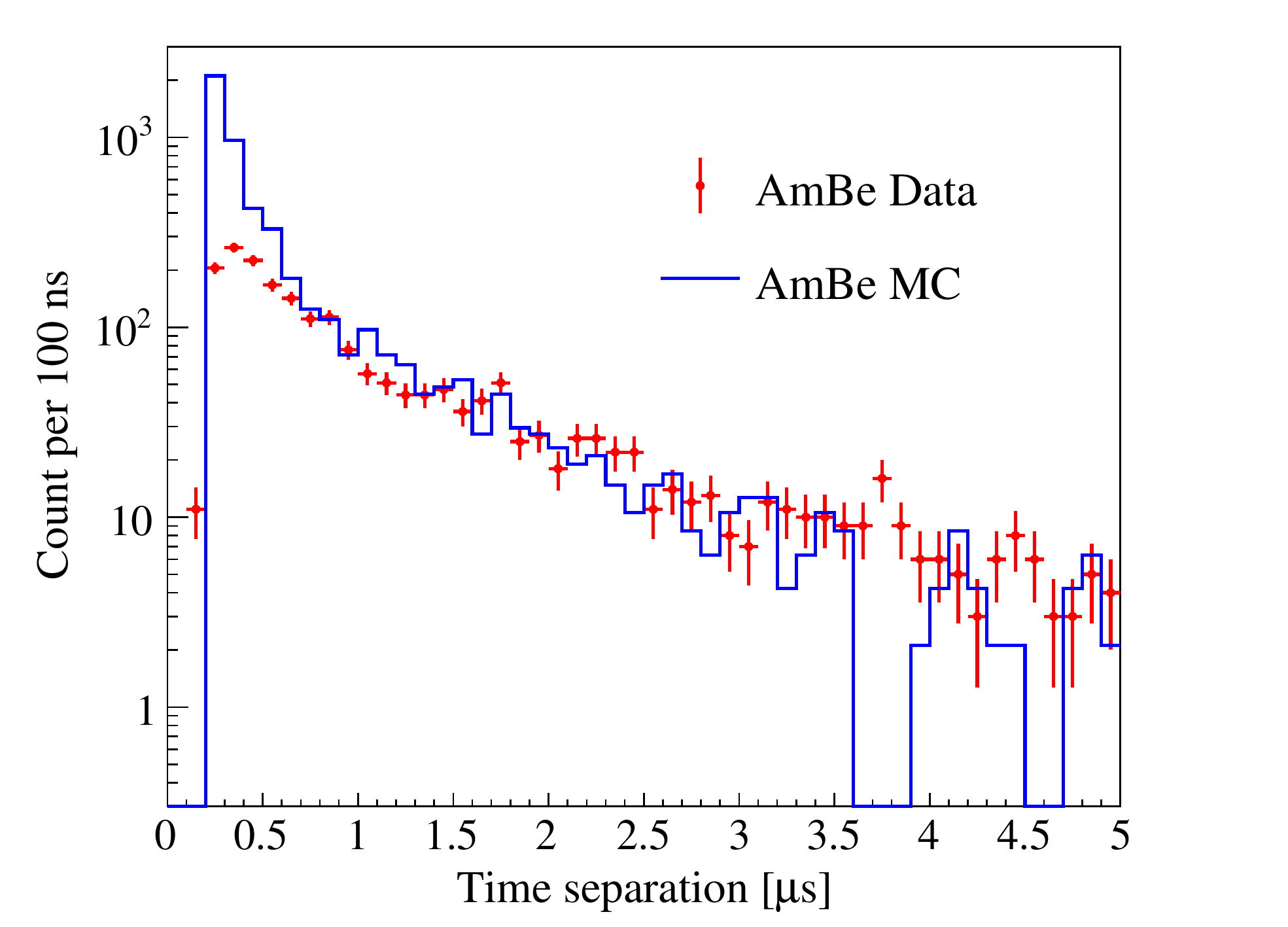}
  \caption{The time separation between the 40~keV prompt $\gamma$ rays
  from $^{129}$Xe$^*$ and the delayed HEGs from neutron capture in Run~10 AmBe data (red) and MC (blue).}
  \label{fig:time_separation_comparison}
\end{figure}

To validate the neutron capture process in the MC, we compare the time
separation between the prompt and delayed energy deposition in the
data and MC. Many neutron events in the data do not have a prompt $S1$
because the prompt energy is deposited in the dead region. To cleanly
select events with a genuine prompt $S1$, we make a cut at 40~keV in
both the data and MC. This energy corresponds to the de-excitation $\gamma$
of $^{129}$Xe excited by the neutron. 
Then we look for the second $S1$ corresponding to the
delayed capture of the same neutron and obtain the distribution of the time difference . A
comparison between the data and MC is shown in
\cref{fig:time_separation_comparison}, where one observes a fairly
good agreement above 1~$\mu$s. The discrepancy below 1~$\mu$s is
likely due to the inefficiency in identifying two $S1$s from the data
when they are close-by. On the other hand, the selection of HEGs from
the data does not require prompt-delayed S1 pairs, therefore is entirely unaffected.

Lastly, to quantify the global agreement between the AmBe data and MC,
we compare the ratios between SSNR and HEG events in
\cref{tab:AmBe_lowe_highe_ratio}. For both run periods, the data
and MC ratios are consistent within 4\%. An increased number of SSNR events
by a factor of 20\% is obtained by an MC simulation without the 4.4~MeV
$\gamma$ in the source generator.
The difference would further increase when the isolated neutrons are
produced at the PTFE or PMTs closer to the xenon target. This demonstrates
the importance of correlated $\gamma$ emission in the neutron
production model.

\section{The improved simulation}
\label{sec:material_neutron}
In this section, we developed an improved MC to calculate
 the ratio
between SSNR and HEG events induced by radiogenic neutrons ($R_{\rm{mc}}$ in Eqn.~\ref{eq:heg_rmc}) in the DM data.

\subsection{Evaluation of neutron production rate}
$^{238}$U, $^{235}$U, and $^{232}$Th decay chains are long-lived $\alpha$ emitters.
Due to the low Columb potential, ($\alpha$,n) events are mostly 
\linebreak \vspace{-8mm}

\noindent produced on
the light nuclei in the detector, such as fluorine in the PTFE
reflector wall, aluminum in the ceramic stem of the PMTs, silicon in the quartz
window of the PMTs, etc. In addition, those decay chains also produce SF
neutrons. As the input to the simulation, the radioactivities of the
most important components for neutron production in the detector are
summarized in \cref{tab:material_radioactivity}.

\begin{table*}[t]
\footnotesize
\caption{The activities of the materials used for neutron background
    estimation in the PandaX-II experiment.  For $^{238}$U, its early
    chain ($^{238}$U$_{\rm e}$) represents $^{238}$U
    $\to$ $^{230}$Th, and the later chain ($^{238}$U$_{\rm l}$) refers to 
    $^{226}$Ra $\to$ $^{206}$Pb. For $^{232}$Th, the early chain ($^{232}$Th$_{\rm e}$) and later chain ($^{232}$Th$_{\rm l}$) refer to $^{232}$Th $\to$ $^{228}$Ac and $^{228}$Th $\to$ $^{208}$Pb, respectively. The activities of the SS and PTFE were measured with the PandaX counting station~\cite{Wang:2016eud} assuming secular equilibrium. The radioactivities from the 3-in PMTs are taken from Ref.~\cite{Aprile:2015uzo}.}
\label{tab:material_radioactivity}
\doublerulesep 0.1pt \tabcolsep 10pt
\centering
\begin{tabular}{cc|ccccc}
\toprule
    \multirow{2}*{\shortstack{Component (Material)}} & \multirow{2}*{\shortstack{Quantity}} &     \multicolumn{5}{c}{Radioactivity (mBq/kg or mBq/piece)} \\\cline{3-7}
    & & $^{238}$U$_{\rm e}$ & $^{238}$U$_{\rm l}$ & $^{235}$U & $^{232}$Th$_{\rm e}$ & $^{232}$Th$_{\rm l}$\\
    \hline
    PTFE & 55.5 kg & 3.2 & 3.2 &1.3 & 1.5 & 1.5\\
    Cryostat (SS) & 256.9 kg & 1.7 & 1.7 & 2.4 & 2.7 & 2.7\\
    PMT stem (Al$_2$O$_3$) & 110 piece & 2.4 & 0.3 & 0.1 & 0.2 & 0.1\\
    PMT window (SiO$_2$) & 110 piece & 1.2	& 0.07 & 0.02 & 0.03 & 0.03\\
\bottomrule
\end{tabular}
\end{table*}

The neutron yields, $Y_{ij}$, are calculated with SOURCES-4A and summarized in \cref{tab:neutron_yield}. The SF neutron
yield of $^{238}$U$_{\rm e}$ is 1.1 $\times10^{-6}$ neutrons/decay, and
that of other chains are negligible. The neutron production
rate by each component, $\sum_{j}n_{ij}$, by combining the
radioactivities and neutron yields, is also listed in the table, where
dominating contribution from PTFE can be observed.  The typical
contribution to neutron production from SF is about 4\% of that from
($\alpha$,n).

\begin{table*}[t]
\footnotesize
\caption{Neutron yields of $^{238}$U, $^{235}$U, and $^{232}$Th in different materials. Among all channels, fluorine has the
    largest ($\alpha$,n) neutron yield. SF only happens in the earl chain, with a yield of about 2 orders less compared to the ($\alpha$,n). 
    The neutron production rates, $\sum_jn_{ij}$, are also listed for corresponding component $i$.}
\label{tab:neutron_yield}
\tabcolsep 18pt
\centering
\begin{tabular}{c|ccccc|c}
\toprule
    \multirow{2}*{\shortstack{Material}} & \multicolumn{5}{c|}{Neutron yield (neutrons/decay)} & $\sum_jn_{ij}$\\\cline{2-6} 
    & $^{238}$U$_{\rm e}$ & $^{238}$U$_{\rm l}$ & $^{235}$U & $^{232}$Th$_{\rm e}$ & $^{232}$Th$_{\rm l}$ & (neutrons/day) \\
    \hline
    PTFE	& 7.6 $\times10^{-6}$ & 5.5 $\times10^{-5}$ & 8.9 $\times10^{-5}$ & 7.9 $\times10^{-7}$	& 8.6 $\times10^{-5}$ & 2.1 $\times10^{0}$\\
    SS & 1.1 $\times10^{-6}$ & 4.1 $\times10^{-7}$ & 3.4 $\times10^{-7}$ & 4.9 $\times10^{-11}$ & 1.5 $\times10^{-6}$ & 1.7 $\times10^{-1}$\\
    Al$_2$O$_3$ & 1.3 $\times10^{-6}$ & 6.3 $\times10^{-6}$ & 9.5 $\times10^{-6}$ & 1.3 $\times10^{-8}$ & 1.1 $\times10^{-5}$ & 6.6 $\times10^{-2}$ \\
    SiO$_2$ & 1.2 $\times10^{-6}$ & 9.9 $\times10^{-7}$ & 1.5 $\times10^{-6}$ & 7.8 $\times10^{-9}$ & 1.6 $\times10^{-6}$ & 1.5 $\times10^{-2}$\\
\bottomrule
\end{tabular}
\end{table*}

\subsection{New ($\alpha$, n) generators}
\label{sec:neutron_alpha_n}
In recent years, improved low energy nuclear models and data
libraries have been incorporated into Geant4~\cite{Geant4Physics}. Therefore, as a remedy
to SOURCES-4A, we utilized Geant4 (version 10.03p03) and developed a
new ($\alpha$, n) neutron generator to include the excitation/de-excitation of the
outgoing nucleus.

In Geant4, the {\em INCL++} model handles the nuclear reactions and
{\em G4ExcitationHandler} models the nuclear
excitations/de-excitations~\cite{Geant4Physics}. Long-lived decay
chains are launched automatically by Geant4, and for events where
the neutron is produced via ($\alpha$, n), the information of the
energy and particle IDs of the production vertices are
stored. It is identified that in the native {\em
  G4ExcitationHandler}, the first $\gamma$ in a cascade disrespects the
known energy level in the final nucleus.  This causes a significant
and incorrect suppression for the ($\alpha$,n$_0$) channel relative to the
others. A fix is implemented to correct the inappropriate $\gamma$
energy and absorb the difference to the kinetic energy of the
neutron~\footnote{Q. H. Wang, et~al., in preparation (2019).}. 
The corrected kinetic energy of the neutron and
different $\gamma$ energies (if any) are then stored and later used as
the primary generator of BambooMC. For simplicity, the directions
of neutron and $\gamma$(s) are assumed to be isotropic.

We carried out a BambooMC simulation with the
improved generator for $^{238}$U in PTFE. The results are summarized
in \cref{tab:PTFE_U238_gamma_effect}, together with that by the
SOURCES-4A generator. Clearly, the former leads to a suppression  \linebreak \vspace{-8mm}

\begin{table}[H]
\footnotesize
\caption{The $P_{\rm ssnr}$, $P_{\rm heg}$, and their ratio predicted for $^{238}$U in PTFE by the two ($\alpha$,n) generators. 10$^6$ events are generated in both simulations. See text for details.}
\label{tab:PTFE_U238_gamma_effect}
\doublerulesep 0.1pt \tabcolsep 10pt
\centering
\begin{tabular}{cccc}
\toprule
    Generator & $P_{\rm ssnr}$ & $P_{\rm heg}$ & Ratio\\
    \hline
    Improved & 1038/10$^6$ & 27602/10$^6$ & 1/26.6\\
    SOURCES-4A & 3997/10$^6$ & 26171/10$^6$ & 1/6.5\\
\bottomrule
\end{tabular}
\end{table}

\noindent in the SSNR events by a factor of four due to the correlated $\gamma$ emission,
consistent with the expectation in \cref{sec:ambemc}.

\subsection{New SF generator}
Similar to the ($\alpha$, n) reaction, multiple neutrons and $\gamma$s
are produced in the SF of $^{238}$U. The average neutron and $\gamma$
multiplicities are 2.0 and 6.4, respectively~\cite{Verbeke:2016},
which was not taken into account in the SOURCES-4A generator. Following
the practice in Ref.~\cite{Verbeke:2016}, we employed a special
Geant4-based program to simulate the SF of $^{238}$U. This program 
incorporates the LLNL Fission Library 2.0.2 including a special FREYA
model~\cite{Verbeke:freya} that takes into account the correlations
in neutron multiplicity, energy, angles, and the energy sharing
between neutrons and $\gamma$s. 
Using this new generator with BambooMC, we generated SF
of $^{238}$U in the PTFE materials, and the resulting SSNR and HEG events
are shown in \cref{tab:PTFE_U238SF_ng_effect}. Also shown in the
same table are those using the SOURCES-4A generator. Due to the large
neutron and $\gamma$ multiplicity, the SSNR events are suppressed to a
negligible level in comparison to the ($\alpha$, n) reaction.

\begin{table}[H]
\footnotesize
\caption{The $P_{\rm ssnr}$, $P_{\rm heg}$, and their ratio predicted for $^{238}$U in PTFE by the two SF generators. 10$^6$ events are generated in both simulations. ``2.0'' 
is the mean number of neutrons per fission. See text for details.}
\label{tab:PTFE_U238SF_ng_effect}
\doublerulesep 0.1pt \tabcolsep 10pt
\centering
\begin{tabular}{cccc}
\toprule
    Generator & $P_{\rm ssnr}$ & $P_{\rm heg}$ & Ratio\\
    \hline
    Improved & 89/2.0/10$^6$ & 86693/2.0/10$^6$ & 1/974.1\\
    SOURCES-4A & 3885/10$^6$ & 27771/10$^6$ & 1/7.1\\
\toprule
\end{tabular}
\end{table}

\subsection{Evaluation of $R_{\rm{mc}}$}
\label{sec:r_mc}
We performed BambooMC simulation for all detector materials in
\cref{tab:material_radioactivity} with the improved neutron
generators. The same cuts used in the analysis of AmBe MC were used to
select the SSNR and HEG events. The results are shown in
\cref{tab:material_lowe_highe_ratio}. Due to the high neutron
yield and the high $^{238}$U impurity level, the contribution from
PTFE dominates the SSNR and HEG events. The naively estimated SSNR
background according to Eqn.~\ref{eq:ni} is about 0.2 events for both Run 9 and Run 10,
respectively. As we have mentioned in \cref{sec:method}, these values heavily rely on the input of the radioactivities, therefore lack robustness. The overall
ratios between the SSNR and HEG events are 1/26.6 for Run 9 and 1/24.6
for Run 10. The difference mainly arises from different FV for SSNR
selections in Run 9 and Run 10~\cite{Cui:2017nnn} (due to different
levels of $^{127}$Xe background).  The uncertainties of $R_{\rm mc}$
are estimated to be 16.3\% and 18.3\% for Run 9 and Run 10,
respectively, based on the spectra difference of HEG events in AmBe
data and MC, discussed in \cref{sec:ambemc}.

\begin{table*}[t]
\footnotesize
  \caption{Predicted SSNR and HEG events in counts/day and their ratios in PandaX-II using the new generators.}
  \label{tab:material_lowe_highe_ratio}
  \tabcolsep 10pt
  \centering
  \begin{tabular}{c|ccc|ccc}
\toprule
    \multirow{2}*{\shortstack{Components}} & \multicolumn{3}{c|}{Run 9}  & \multicolumn{3}{c}{Run 10} \\
    \cline{2-7}
    & \# SSNR & \# HEG & Ratio & \# SSNR & \# HEG & Ratio \\
    \hline
    PTFE & $1.9\times10^{-3}$  & $5.8\times10^{-2}$  & 1/30.8  & $2.0\times10^{-3}$ & $5.8\times10^{-2}$ &  1/28.8 \\
    Cryostat & $3.4\times10^{-4}$  & $2.8\times10^{-3}$  & 1/8.2  & $3.5\times10^{-4}$ &  $2.8\times10^{-3}$ & 1/7.9  \\
    PMT Stem & $1.2\times10^{-4}$  & $1.3\times10^{-3}$  & 1/10.4  & $1.6\times10^{-4}$ & $1.3\times10^{-3}$ & 1/7.8 \\
    PMT Window & $5.1\times10^{-6}$  & $3.9\times10^{-4}$  & 1/76.3  & $6.5\times10^{-6}$ & $3.9\times10^{-4}$ &  1/59.7\\
    \hline
    Total & $2.3\times10^{-3}$ & $6.2\times10^{-2}$  & 1/26.6  &  $2.6\times10^{-3}$&  $6.3\times10^{-2}$& 1/24.6 \\
\bottomrule
  \end{tabular}
\end{table*}

\section{Improved neutron background estimation in PandaX-II}
\label{sec:pandax_nbkg}
As discussed in Secs.~\ref{sec:ambe_data} and \ref{sec:ambemc},
neutrons can induce HEG events in xenon target, with a rate much higher than
the SSNR events. Therefore in PandaX-II DM data, we look for HEGs and
estimate the corresponding SSNR background from detector materials
using the $R_{\rm mc}$ in \cref{sec:r_mc}.

\begin{figure}[H]
  \centering
  \includegraphics[width=1.0\columnwidth]{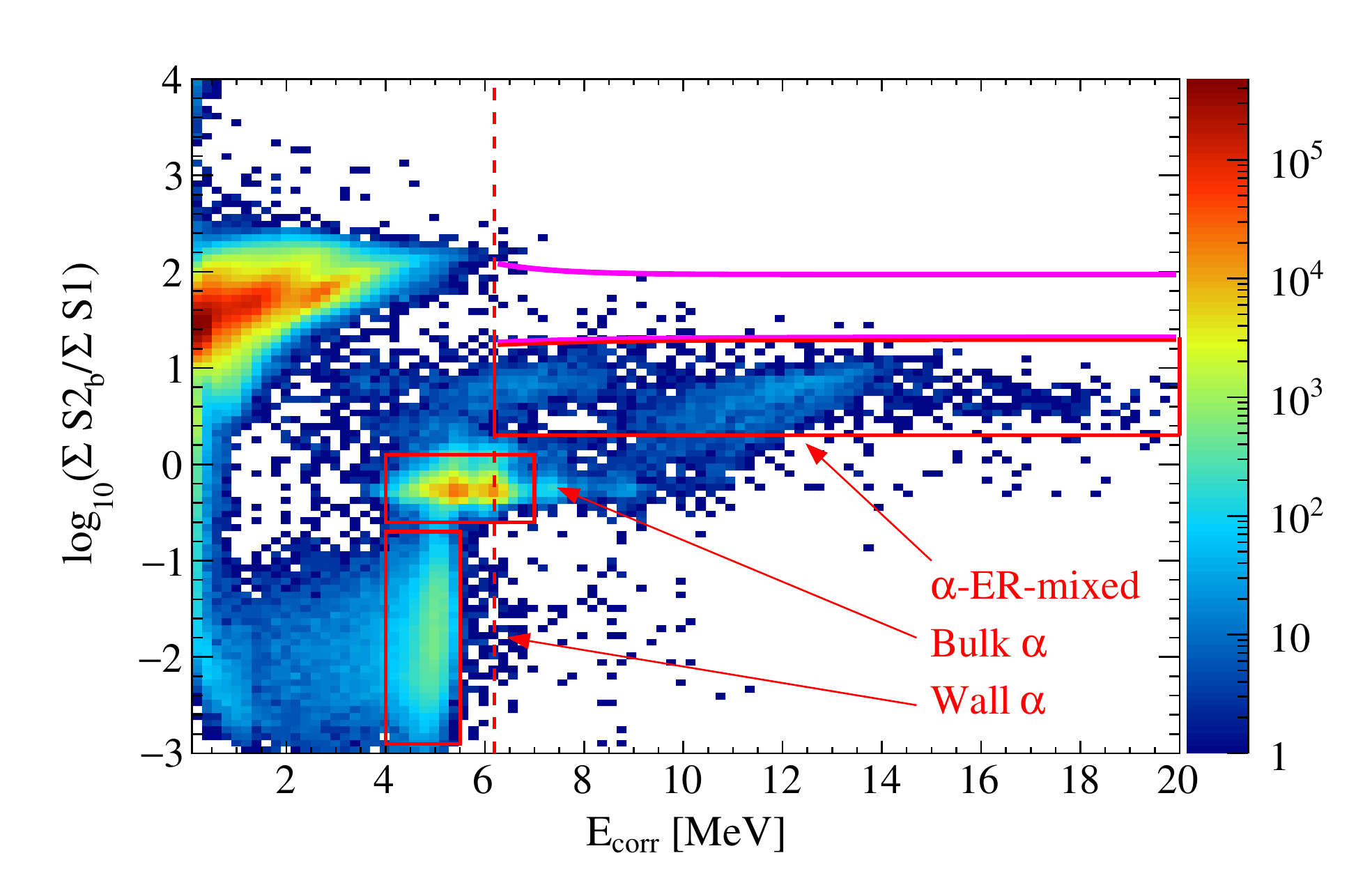}
  \includegraphics[width=1.0\columnwidth]{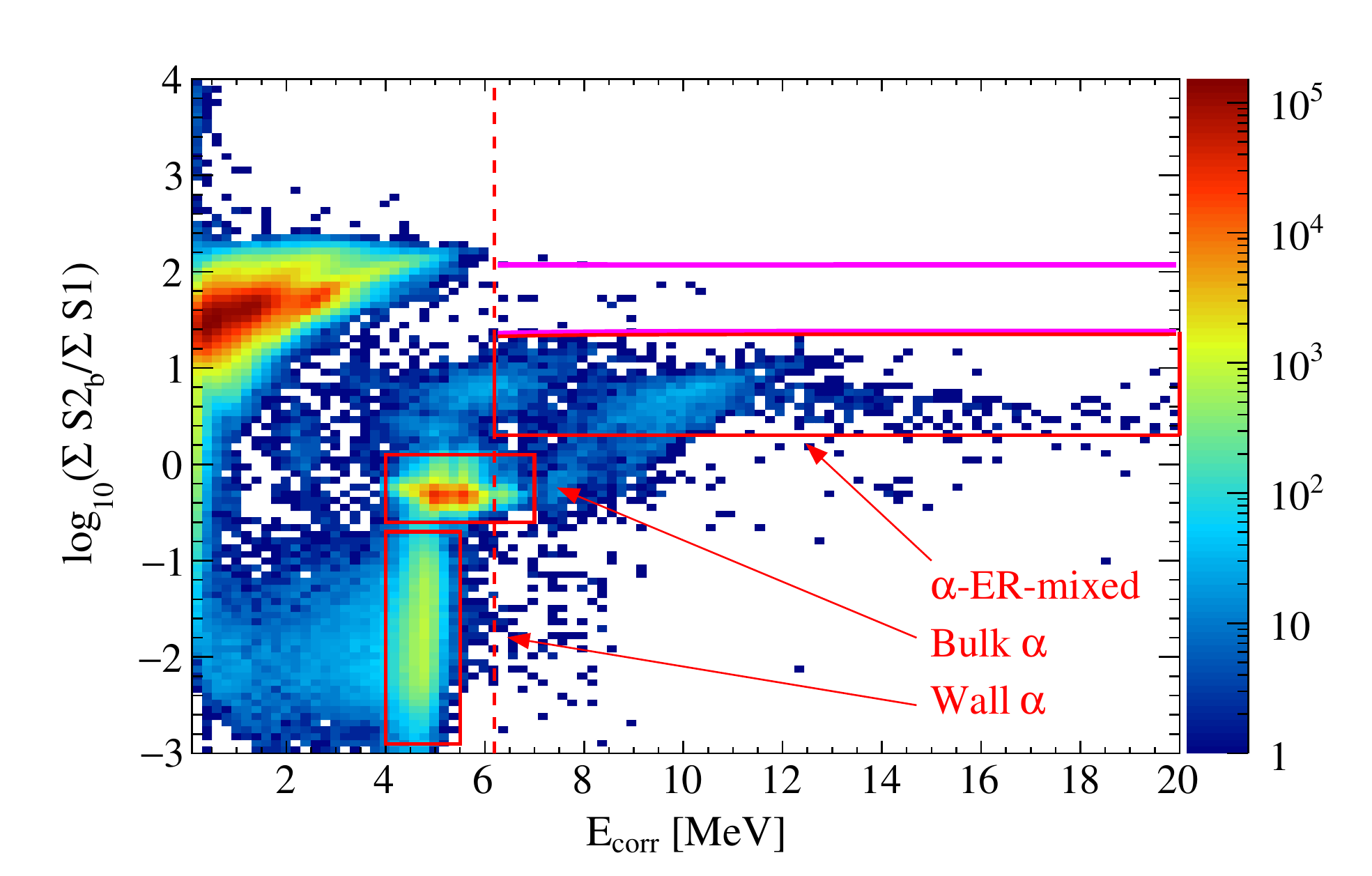}
  \caption{The distributions of events in $\log_{10}(\sum S2_{b}/\sum
    S1)$ vs. $E_{\rm corr}$ in the Run 9 (top) and Run 10 (bottom) DM data.}
  \label{fig:DM_energy_band}
\end{figure}

We selected HEGs from the DM data using the same procedures
as those in analyzing the AmBe data. At the high energy region
of a few MeVs, the distributions of $\log_{10}(\sum S2_{b}/\sum S1)$
vs. $E_{\rm{corr}}$ in Run 9 and Run 10 DM data are shown in
\cref{fig:DM_energy_band}. The fact that the PMTs in Run 9 and
Run 10 were operated at different high voltage~\cite{Cui:2017nnn}
leads to different saturation for large $S1$s and $S2$s, thereby some
difference in the high energy event distributions. 36 and 20 HEG
candidates are found in Run 9 and Run 10, respectively. However,
different from the AmBe data, due to the much lower rate of HEGs, the
$\alpha$-ER-mixed events become more pronounced and can contaminate
the true HEGs.

Since an $\alpha$ cannot scatter twice in the detector, one expects
that the $\alpha$-mixed events contain less amount of $S2$s than the
\linebreak \vspace{-8mm}

\begin{figure}[H]
\footnotesize
  \centering
  \includegraphics[width=1.0\columnwidth]{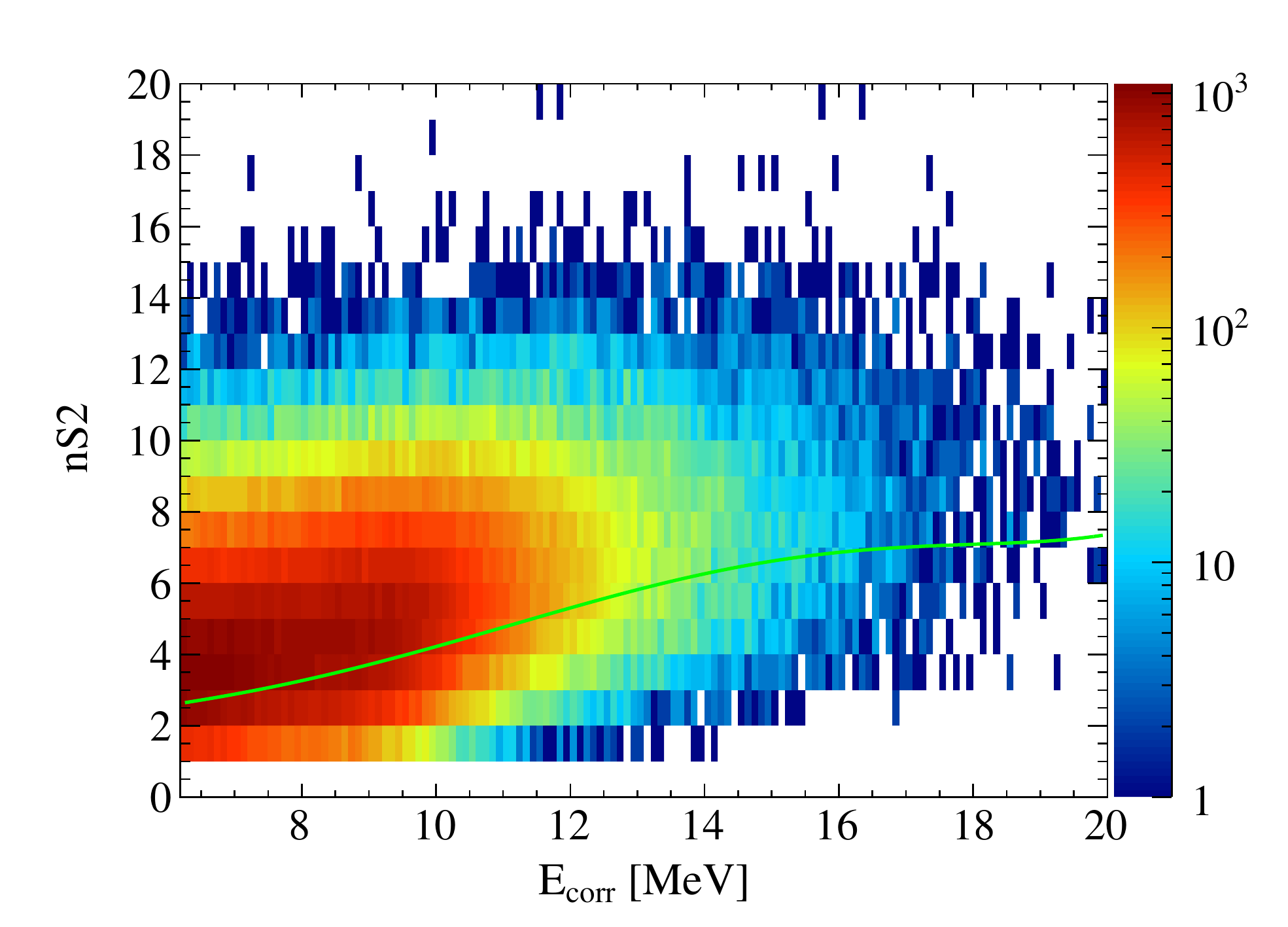}
  \includegraphics[width=1.0\columnwidth]{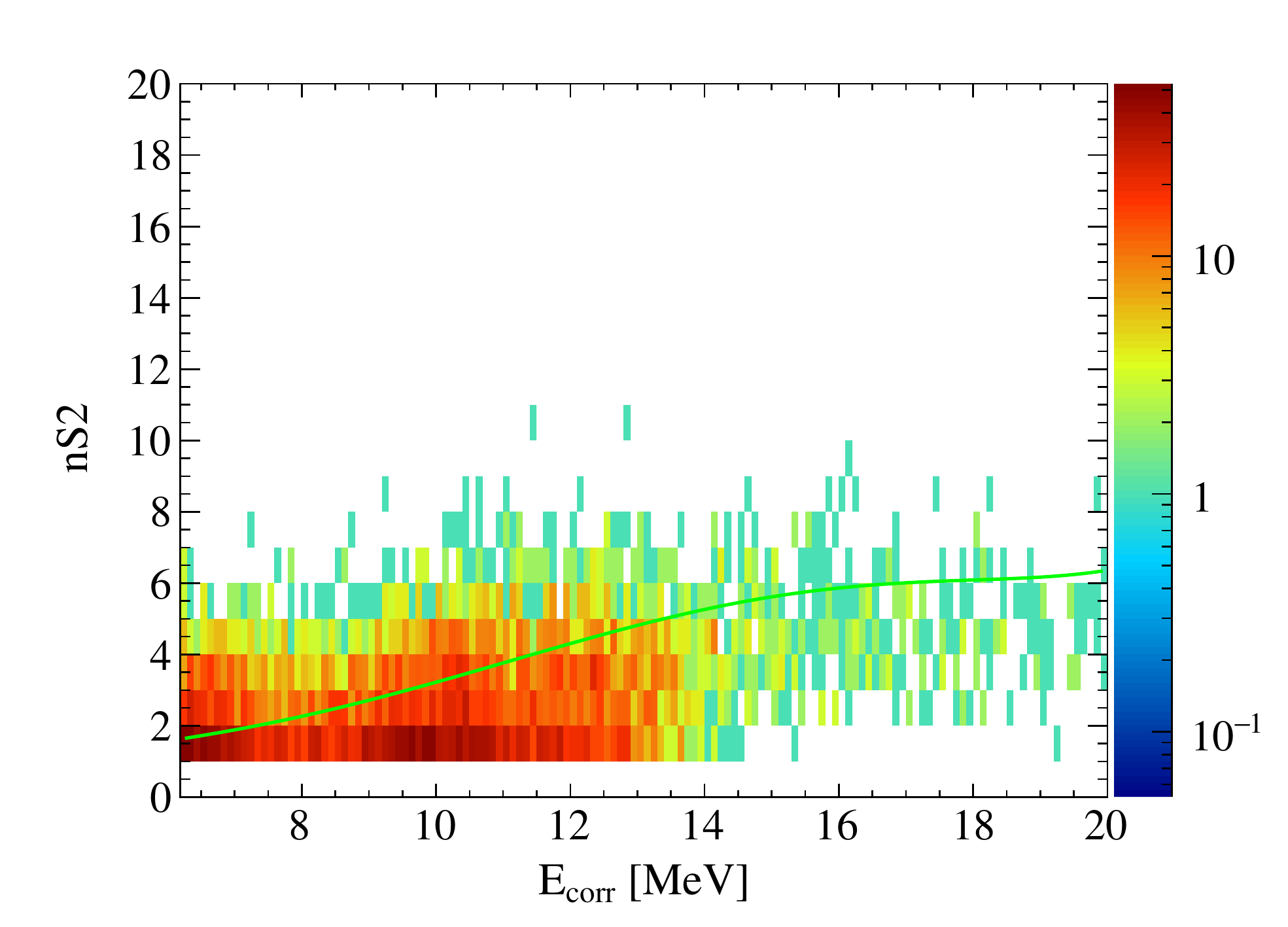}
  \caption{The distribution of number of $S2$s vs. $E_{\rm{corr}}$ for
    HEGs in AmBe data (top) and high energy $\alpha$-mixed events in DM
    data (bottom), Run 9 and Run 10 combined. The green curve represents
    the $\alpha$-rejection cut.}
  \label{fig:alpha_rejection_cut}
\end{figure}

\noindent $\gamma$ cascade in the HEGs. This is confirmed further in
\cref{fig:alpha_rejection_cut}, where the distributions of the
number of $S2$s vs. $E_{\rm corr}$ for HEGs in AmBe data and the 
$\alpha$-mixed events below the ER band in DM data 
(as indicated in \cref{fig:DM_energy_band})
are compared.  An $\alpha$-rejection cut is developed (green curve in
\cref{fig:alpha_rejection_cut}), with an 82\% (61\%) efficiency on
HEGs for Run 9 (Run 10) estimated from AmBe data, with an estimated 
systematic uncertainty of 10.5\%. The corresponding
rejection power for $\alpha$-mixed events are 91$\pm$1\% and 89$\pm$1\%,
respectively, estimated from DM data.

\begin{figure}[H]
  \centering
  \includegraphics[width=1.0\columnwidth]{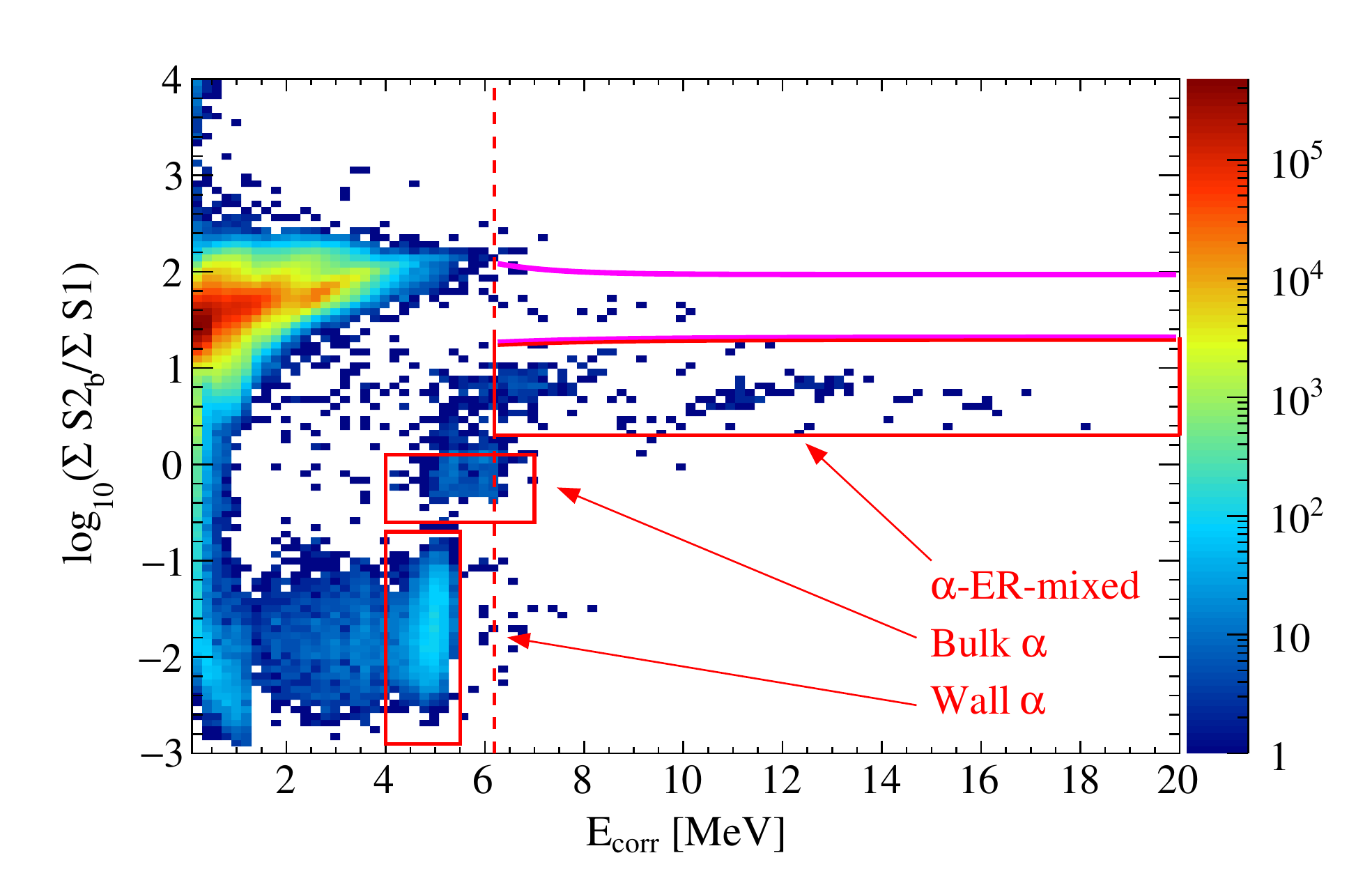}
  \includegraphics[width=1.0\columnwidth]{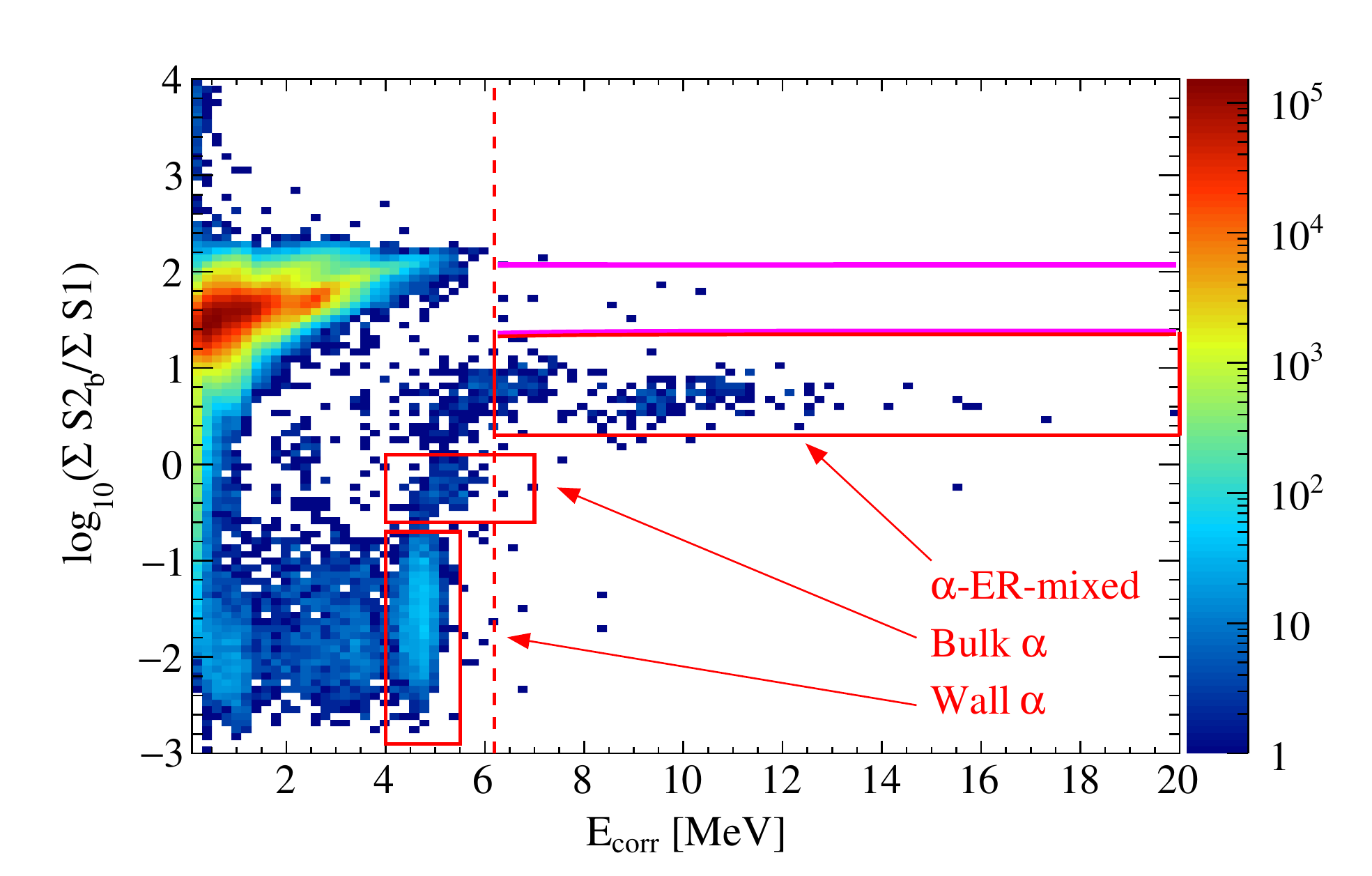}
  \caption{The distribution of $\log_{10}(\sum S2_{b}/\sum S1)$ vs. $E_{\rm corr}$ for
    events with the $\alpha$-rejection cut in Run 9 (top) and Run 10 (bottom) DM data.}
  \label{fig:DM_energy_band_with_alpha_rejection_cut}
\end{figure}

\begin{figure}[H]
  \centering
  \includegraphics[width=1.0\columnwidth]{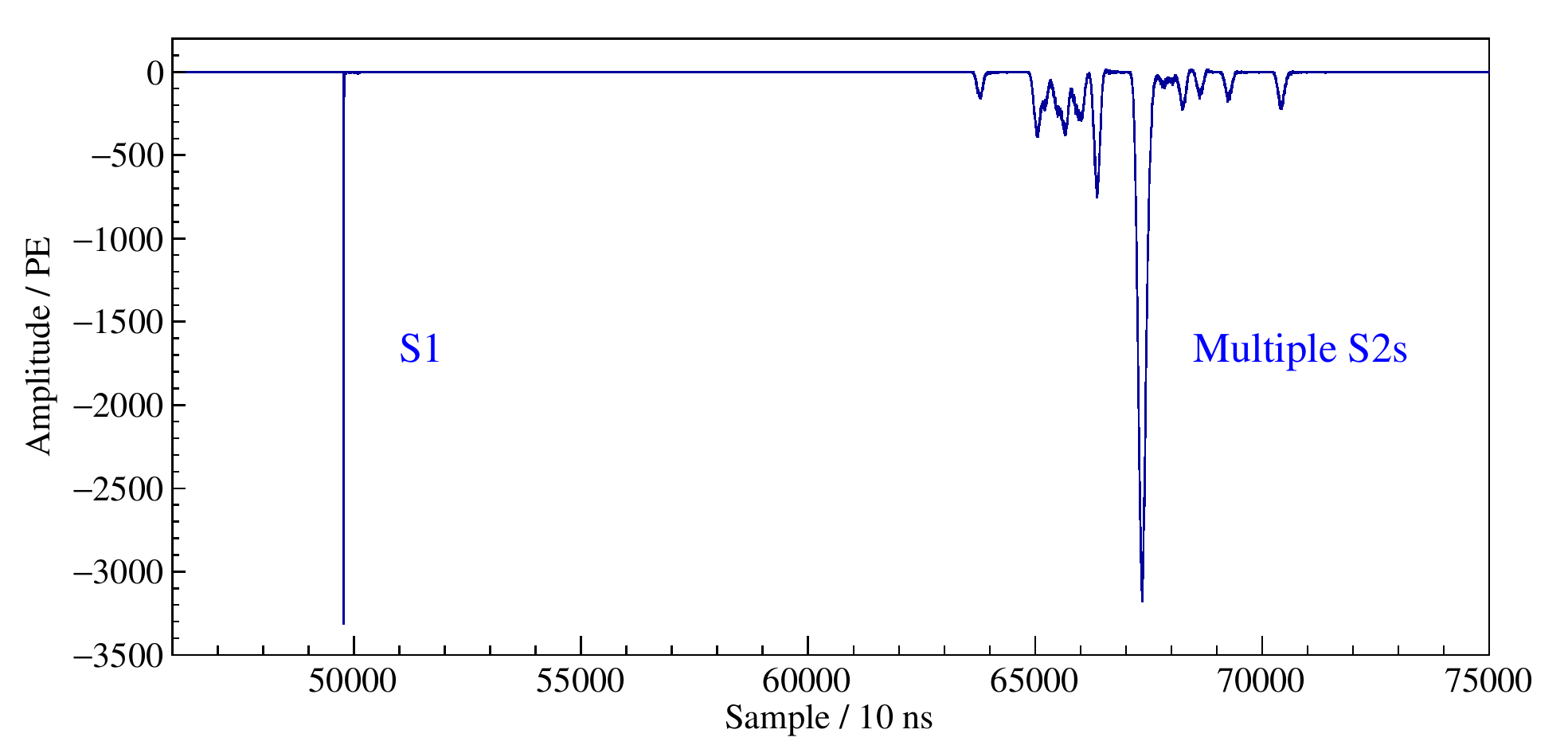}
  \caption{An example HEG waveform from Run 10 DM data.}
  \label{fig:heg_waveform_dm}
\end{figure}

After applying the $\alpha$-rejection cut, the band distributions of
the surviving events in Run 9 and Run 10 DM data are shown in
\cref{fig:DM_energy_band_with_alpha_rejection_cut}. In total, 16
and 8 HEG candidates survive in Run 9 and Run 10, respectively. The
final HEG candidates have multiple $S2$s, consistent with those from
the AmBe data, with an example waveform shown in
\cref{fig:heg_waveform_dm}. The residual $\alpha$-mixed background
is estimated to be 1.7$\pm$0.7 and 0.9$\pm$0.7 events based on the cut
efficiency and rejection power mentioned above. The background
subtracted and efficiency corrected HEGs, $N_{\rm heg}$, are 
17.5$\pm$5.6 events in Run 9 and 11.6$\pm$5.7 events in Run 10, where 
the statistical and systematic uncertainties are combined.

Using $N_{\rm heg}$ obtained from the data and Eqn.~\ref{eq:heg_rmc},
the improved estimation of SSNR background is now 0.66$\pm$0.24 and
0.47$\pm$0.25 events for Run 9 and Run 10, respectively, in which the
uncertainties of $N_{\rm heg}$ and $R_{\rm mc}$ are combined.  The 
final results are summarized in \cref{tab:pandax_ii_nbkg},
together with the estimated neutron background from our previous
publication~\cite{Cui:2017nnn}. This new evaluation presents a major improvement in the SSNR probability calculation and an {\it in situ} normalization against the measured HEGs in the data, therefore supersedes the previous.

\begin{table}[H]
\footnotesize
  \caption{The neutron background events in Run 9 and Run 10 evaluated with the improved method. The results from our previous publications are shown for comparison.}
  \label{tab:pandax_ii_nbkg}
  \doublerulesep 0.1pt \tabcolsep 10pt
  \centering
  \begin{tabular}{ccc}
    \hline\hline
    Methods & Run 9 DM & Run 10 DM\\
    \hline
    This work & 0.66 $\pm$ 0.24 & 0.47 $\pm$ 0.25\\
    Ref.~\cite{Cui:2017nnn} & 0.85 $\pm$ 0.43 & 0.83 $\pm$ 0.42 \\
    \hline\hline
  \end{tabular}
\end{table}

\section{Summary}
\label{sec:summary}
We performed a detailed study on the neutron background originating
from detector materials in the PandaX-II experiment. An improved 
neutron generator model that incorporates the correlated emission of
neutron(s) and $\gamma$(s) was developed and benchmarked. A novel 
data-driven method, with a robust connection between the SSNR and HEG
events, was used to predict the neutron background in PandaX-II. The
improved neutron background levels are lower than those reported
before and are more reliable with well-controlled uncertainties.
In the next generation of 
PandaX-II, i.e., the 4-ton scale PandaX-4T experiment~\cite{Zhang:2018xdp},
a solid estimation of the neutron background is expected to be 
achieved by applying this new method. It can be more generally employed
in the neutron background evaluation in other dark matter experiments.

\Acknowledgements{
This project has been supported by a Double Top-class grant from Shanghai Jiao
Tong University, grants from National Science Foundation of China
(Nos. 11435008, 11505112, 11525522, 11775141 and 11755001),
a grant from the Ministry of Science and Technology of China
(No. 2016YFA0400301). We thank the Office of Science and Technology,
Shanghai Municipal Government (No. 11DZ2260700, No. 16DZ2260200,
No. 18JC1410200) and the Key Laboratory for Particle Physics,
Astrophysics and Cosmology, Ministry of Education, for important
support. This work is supported in part by the Chinese Academy of
Sciences Center for Excellence in Particle Physics (CCEPP) and Hongwen
Foundation in Hong Kong. Finally, we thank the CJPL administration and
the Yalong River Hydropower Development Company Ltd. for indispensable
logistical support and other help.}

\end{multicols}
\end{document}